\documentclass[aps,prl,twocolumn,showpacs,floatfix,superscriptaddress]{revtex4}

\usepackage{amsmath,amssymb}
\usepackage{graphicx,color}
\usepackage{times}

\newcommand{\varH}{{\mathcal{H}}}
\newcommand{\bfk}{{\mathbf{k}}}
\newcommand{\bfq}{{\mathbf{q}}}

\newcommand{\up}{{\uparrow}}
\newcommand{\down}{{\downarrow}}

\let\up=\uparrow%
\let\down=\downarrow%

\begin{document}

\title{ Josephson current in strongly correlated double quantum dots}

\author{Rok \v{Z}itko}
\affiliation{J. Stefan Institute, Jamova 39, SI-1000 Ljubljana, Slovenia}

\author{Minchul Lee}
\affiliation{Department of Physics, Kyung Hee University, Yongin 446-701, Korea}

\author{Rosa L\'opez}
\affiliation{Departament de F\'{i}sica, Universitat de les Illes Balears,
  E-07122 Palma de Mallorca, Spain}

\author{Ram\'on Aguado}
\affiliation{Teor\'{\i}a de la Materia Condensada, Instituto de Ciencia de
  Materiales de Madrid (CSIC) Cantoblanco,28049 Madrid, Spain}

\author{Mahn-Soo Choi}
\affiliation{Department of Physics, Korea University, Seoul 136-701, Korea}

\date{\today}

\begin{abstract}
We study the transport properties of a serial double quantum dot (DQD) coupled to
two superconducting leads, focusing on the Josephson current through the DQD
and the associated $0$-$\pi$ transitions which result from the subtle interplay
between the superconductivity, the Kondo physics, and the inter-dot superexchange
interaction.  We examine the competition between the superconductivity and the Kondo
physics by tuning the relative strength $\Delta/T_K$ of the superconducting
gap $\Delta$ and the Kondo temperature $T_K$, for different strengths
of the superexchange coupling determined by the interdot tunneling $t$ relative to the dot level
broadening $\Gamma$.  We find strong renormalization of $t$, a significant role
of the superexchange coupling $J$, and a rich phase diagram of the
$0$ and $\pi$-junction regimes.  In particular, when both the superconductivity and the
exchange interaction are in close competion with the Kondo physics ($\Delta\sim J\sim
T_K$), there appears an island of $\pi'$-phase at large values of the
superconducting phase difference.
\end{abstract}

\pacs{73.23.-b,72.15.Qm,74.45.+c} \maketitle
\begin{figure}[!b]
\centering \includegraphics[width=8cm]{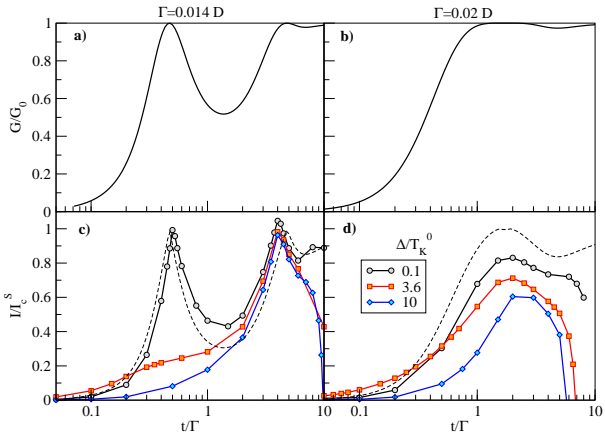} 
\caption{(color online) Normal state conductance (a,b) and the
critical Josephson current in the superconducting state (c,d) as a
function of $t/\Gamma$ for $\Gamma=0.014D$ (a,c) and $\Gamma=0.02D$
(b,d). The results are expressed in units of the conductance quantum
$G_0=2e^2/h$ and the supercurrent quantum $I_c^s=e\Delta/\hbar$. In
the superconducting case (c,d), the different curves are for
$\Delta/T_K=0.1$ (black circle), $3.6$ (red square) , and $10$ (blue
diamond).  The dashed line is from an effective non-interacting
theory, Eq.~(\ref{superdots::eq:1}).
  }
\label{superdots::fig:1}
\end{figure}
\begin{figure}
\centering
\includegraphics[width=8cm]{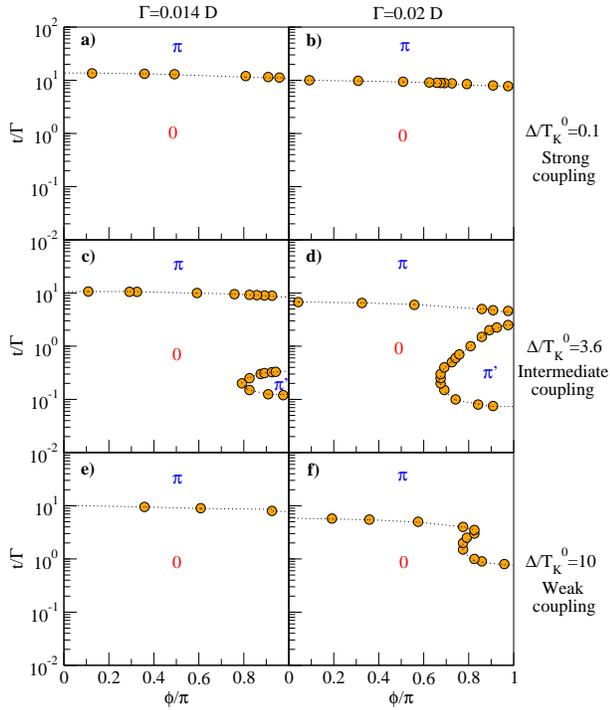}
\caption{Phase boundaries between the $0$- and $\pi$-states of
  the Josephson current for $\Gamma=0.014D$ (left panels) and $\Gamma=0.02D$
  (right panels).  When both the inter-dot superexchange coupling and the
  superconductivity are in close competition with the Kondo effect between the
  superconducting leads and adjacent dots ($t/\Gamma\sim0.2$, $\Delta\sim
  T_K$), there appears an island of $\pi'$-phase at larger phase difference.}
\label{superdots::fig:2}
\end{figure}

\paragraph{Introduction.---}
In a metal containing a dilute concentration of magnetic impurities the competititon between
Kondo physics, which favours screening of the localized spins
by the itinerant conduction-band electrons~\cite{hew93}, and
antiferromagnetic (AF) exchange interactions between impurities leads
to a quantum phase transition~\cite{Jones88}.  Even more interesting
properties emerge when the metal turns superconducting below the
critical temperature. For s-wave superconductors, Cooper pairs formed
by itinerant electrons~\cite{Tinkham96a} are yet another possible
singlet state which competes with the above. The intriguing interplay
of these phenomena, which might actually coexist in complex materials
such as heavy-fermion superconductors, governs the low temperature
physics of these systems.  Nanoscale systems allow to tune the ratio
between the relevant parameters (the Kondo temperature $T_K$, the AF
exchange interaction $J$, and the superconducting gap $\Delta$,
respectively) and, therefore, enable thorough investigations of such
competition in a controlled setting. In the simplest case of single
quantum dots attached to superconducting reservoirs, where only Kondo
physics and superconductivity are relevant, a sign change of the
Josephson current, from positive $0$-junction to negative
$\pi$-junction behavior, signals a quantum phase transition between a
singlet and a doublet ground state as $T_K/\Delta$
decreases~\cite{super-magnetic,ChoiMS04a,super-kondo}. This $0$ to
$\pi$-junction transition has been experimentally realized, confirming
some of these physical aspects~\cite{expsuper}. A double quantum dot
(DQD) coupled to normal metals constitutes a physical realization of
the two-impurity Kondo model~\cite{Georges99a,dqdkondo, Minchul09}, as
demonstrated experimentally~\cite{chan,Craig}. When the reservoirs
become superconducting, this system is a minimal artificial
realization of the described competition among three different
spin-singlet ground states.  
\begin{figure}[!b]
\centering \includegraphics[width=8cm]{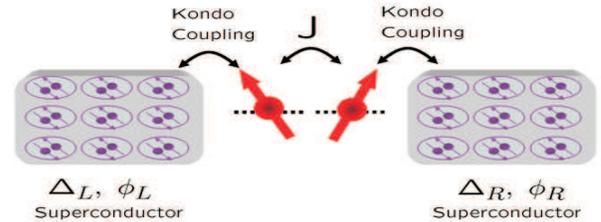} 
\caption{(color online) Schematics of the DQD system coupled to superconducting leads. In the deep Kondo limit, this system is an artificial realization of the two-impurity Kondo problem in the presence of superconducting correlations.
  }
\label{superdots::fig:scheme}
\end{figure}
In this Letter we focus on a detailed analysis of the Josephson
current which, as a ground state property, shows signatures of this
subtle competition.  Our results, obtained by a highly efficient
numerical renormalization group (NRG) scheme, able to deal with
superconducting correlations, are summarized in
Figs.~\ref{superdots::fig:1} and \ref{superdots::fig:2}. The interplay
between superconductivity and the Kondo physics is studied by tuning
the relative strength $\Delta/T_K$.  The role of the superexchange
coupling is tuned by the interdot tunneling $t$ relative to the dot
level broadening $\Gamma$.  
Importantly, we find strong renormalizations of $t$ and a significant
role of the superexchange coupling $J$ compared with the previous
works\cite{Georges99a,dqdkondo,Bergeret06a} based on the slave-boson
mean-field theory (SBMFT).  Moreover, we find a rich phase diagram of
the $0$-$\pi$ transition.  In particular, when both the
superconductivity and the superexchange are in close competion with
the Kondo physics ($\Delta\sim J\sim T_K$), there appears an
unexpected island of $\pi'$-phase at large values of the
superconducting phase difference $\phi$.
We provide clear interpretations by examining the \emph{spin-state-resolved}
Andreev bound states inside the superconducting gap.


\paragraph{Model.---} 
The system that we consider is a DQD modelled as a two-impurity
Anderson model connected to two superconducting leads (Fig.~\ref{superdots::fig:scheme}) described by standard
BCS Hamiltonians~\cite{Tinkham96a}: $\varH = \varH_{\rm D} + \varH_{\rm L} +
\varH_{\rm T}$, where
\begin{align}
  \varH_{\rm D}
  & =
  \sum_i
  \left(
    \epsilon n_i + U n_{i\up} n_{i\down}
  \right)
  - t \sum_\mu \left(d_{1\mu}^\dag d_{2\mu} + (h.c.)\right)
  \\
  \varH_{\rm L}
  & =
  \sum_{\ell\bfk}
  \left[
    \epsilon_{\bfk} n_{\ell\bfk}
    -
    \left(
      \Delta_\ell e^{i\phi_\ell} c_{\ell\bfk\up}^\dag c_{\ell-\bfk\down}^\dag
      + (h.c.)
    \right)
  \right]
  \\
  \varH_{\rm T}
  & =
  V \sum_{\ell\bfk\mu}
  \left[c_{\ell\bfk\mu}^\dag d_{\ell\mu} + (h.c.)\right].
\end{align}
Here $c_{\ell\bfk\mu}$ describes an electron with energy $\epsilon_{\bfk}$,
momentum $\bfk$, and spin $\mu$ on the lead $\ell=1,2$, and $d_{i\mu}$ an
electron in the dot $i=1,2$; $n_{\ell\bfk} \equiv \sum_\mu
c_{\ell\bfk\mu}^\dag c_{\ell\bfk\mu}$ and $n_i \equiv \sum_\mu d_{i\mu}^\dag
d_{i\mu}$.
$\epsilon$ is the single-particle energy on each dot that is tuned by gate
voltages and $U$ is the on-site Coulomb interaction.
The electrons can tunnel between the two dots with the tunneling amplitude
$t$. $\Delta_\ell$ is the superconducting gap, and $\phi_\ell$ the phase
of the superconducting order parameter.

The two leads are assumed to be identical except for the superconducting
phases.
Assuming that the leads have a flat band with the density of states
$\rho=1/2D$, where $2D$ is the bandwidth, the hybridization between the dots
and the leads is well characterized by a single parameter $\Gamma = \pi\rho
V^2$.
Since we are interested in the Kondo correlations, we concentrate on the Kondo
regime with localized level $-\epsilon \gg \Gamma$ and large charging energy
$U \ge 2|\epsilon|$. For the representative results shown below, we choose
$\Gamma=0.014D$ or $\Gamma = 0.02D$, fix $\epsilon=-0.2D$, and take the large
$U=\infty$ limit. The parameter space is nevertheless still large. We examine
the results by varying $\Delta/T_K$, $t/\Gamma$, and $\phi\equiv\phi_L-\phi_R$.

We solve the Hamiltonian using the numerical renormalization group (NRG)
method~\cite{NRG}.
Due to the relatively low symmetry of the problem (in particular due
to the lack of charge conservation), the NRG iteration is numerically
very demanding and special attention is necessary to obtain reliable
results. Using a new discretization scheme the numerical artifacts due
to a large discretization parameter $\Lambda$ are almost completely
canceled out; the technical details are given elsewhere.

We first review two crucial effects from a previous work on DQD with
normal leads \cite{Minchul09} which keep playing significant roles in
the present superconducting case.  Firstly, the inter-dot tunneling
$t$ is significantly renormalized compared with the predictions based
on the SBMFT \cite{Georges99a,dqdkondo,Bergeret06a}. It is important
to treat this effect properly to fully account for the transport.
Secondly, the inter-dot antiferromagnetic superexchange $J$ is finite
even for $U=\infty$.\cite{Minchul09} This interaction is mediated by
the virtual tunneling of electrons to the conduction leads.  It is
thus important to take into account the interplay of the superexchange
coupling with the superconductivity and the Kondo effect.

\paragraph{Strong coupling limit ($T_K\gg\Delta$).---}

Figure~\ref{superdots::fig:1} shows normal state conductance (a,b) and the
critical Josephson current (c,d). Here we focus on the strong
coupling limit, $\Delta/T_K=0.1$ (black circles).
In this limit, the critical Josephson current $I$ shows similar
features as the normal-state conductance. It peaks at equal values
of $t/\Gamma$ and, remarkably, when the conductance in the normal
state is unitary, the Josephson current reaches the quantum limit
$I_c^s=e\Delta/\hbar$.
This is expected since the
Kondo effect dominates over the superconductivity, therefore the
transport is determined by the competition between the Kondo physics
and the interdot superexchange (for $t/\Gamma<5$) or interdot
molecular orbitals ($t/\Gamma>5$).  As we analyzed in detail in the
previous work for the normal lead case, the peaks in $G$ and
$I_c$ at $t/\Gamma\approx 0.4$ (for $\Gamma=0.014D$) result from the
crossover from the ``Kondo singlet'' to the ``superexchange singlet'':
For $t/\Gamma<0.4$, $J<T_K$, wheares for $0.4<t/\Gamma<5$, $J>T_K$.
We stress that the crossover is shifted significantly to smaller
$t/\Gamma=0.4$ compared with the estimation from
SBMFT,\cite{Georges99a,dqdkondo,Bergeret06a} due to the previously
mentioned strong renormalization of $t$.
As $t/\Gamma$ increases beyond $5$, the DQD starts to form molecular
orbitals and effectively behaves as a single QD.
In this regime, the Josephson current in single dots has $\pi$ shift
in the Coulomb blockade
regime~\cite{super-magnetic,ChoiMS04a,super-kondo}, which is directly
related to the $\pi$-phase in Fig.~\ref{superdots::fig:2} (a,b) for
large $t/\Gamma$.


For comparison, we have also calculated the 
critical Josephson current as $I=\max_\phi I(\phi)$, where~\cite{Bee92}
\begin{equation}
\label{superdots::eq:1}
\frac{I(\phi)}{I_c^s} = \frac{g}{2} \sin\phi\sqrt{1 - g \sin^2(\phi/2)},
\end{equation}
where $I_c^s = e\Delta/\hbar$ is the critical current of a transparent
single-mode junction, and $g=G/G_0$ is the (dimensionless)
normal-state conductance obtained from a NRG calculation. These
relations are applicable only under the assumption that the QD state
is weakly affected by the superconductivity. As can be seen in
Fig.~\ref{superdots::fig:1}, for $t/\Gamma \lesssim 5$, the result
from the effective theory (dashed lines) and the fulll NRG calculation
show a qualitative agreement.  For larger $t$ ($t/\Gamma>5$), we
enter the single-dot regime and the Kondo effect is
suppressed.  This leads to the deviation of the two results.


\begin{figure}
\centering \includegraphics*[width=8cm]{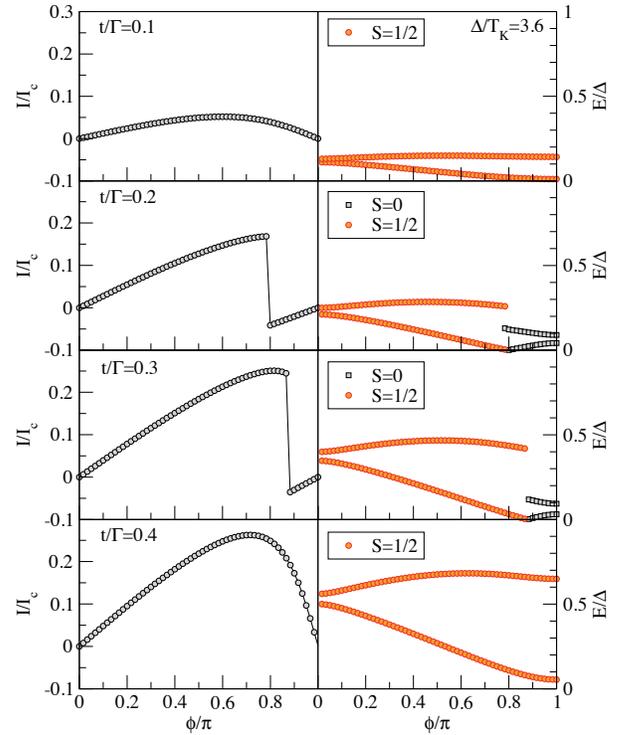} 
\caption{(color on-line) Left panels: Josephson current vs phase
difference near the $\pi'$-phase ($\Delta/T_K=3.6$ and
$t/\Gamma=0.1,0.2,0.3,0.4$, from top to bottom). Right panels:
Corresponding spin-state-resolved Andreev bound states inside the
superconducting gap.  Spin singlet states are depicted by (red)
circles and doublets by (black) squares.  The changes in the spin
states are closely related to the island of $\pi'$ phase in
Fig.~\ref{superdots::fig:2} (c) [similar spin-state-resolved Andreev
states for corresponding parameters will also explain the island in
Fig.~\ref{superdots::fig:2} (d)].} \label{superdots::fig:3}
\end{figure}

\paragraph{Weak coupling limit ($T_K\ll\Delta$).---} The results,
shown in Figs.~\ref{superdots::fig:1}~and~\ref{superdots::fig:2} for
$\Delta/T_K=10$ (gray diamonds, blue online), indicate that the
superconducting correlations in the leads suppress the Kondo effect
and the Josephson current remains small until the system enters the
single-dot regime. It is remarkable that the S-DQD-S system behaves as
a $0$-junction in the \emph{weak coupling limit} in contrast to the
S-QD-S case where the $\pi$-junction appears in the same regime. In
single quantum dots, the appearance of $\pi$-junction is due to the
reversal of the order of the electrons forming Cooper pairs after
tunneling.\cite{super-magnetic}
However, in series coupled DQD, the order is preserved so that no additional
phase factor arises from the tunneling and thus $0$-junction is
formed even in the Coulomb blockade regime.
Specifically, the perturbation theory applied in the weak coupling limit
for $U\to\infty$ gives
\begin{multline}
\frac{I}{I_c^s} =
\sin\phi \sum_{\bfk\bfq}
\frac{2 \Delta t^2 t_\bfk^2 t_\bfq^2}%
{E_\bfk E_\bfq[(\epsilon - E_\bfq)^2 -t^2][(\epsilon - E_\bfk)^2 - t^2]}
\\{}\times
\left(\frac{1}{E_\bfq + E_\bfk} + \frac{1}{2|\epsilon|}\right)
\end{multline}
with $E_\bfk \equiv \sqrt{\epsilon_\bfk^2 + \Delta^2}$.
Our NRG calculations confirm that the ground state is a spin singlet
as long as the $0$-junction is formed. Hence, in contrast to the
single quantum dot system, in a large part of the parameter space
there exists no phase transition as we move from the weak to the
strong coupling limits by varying $\Delta/T_K$: we always have a spin
singlet state with $0$-junction behavior.


Subsequent transition into $\pi$-junction for very large $t/\Gamma$ in
Fig.~\ref{superdots::fig:2} (e,f) is ascribed to the competition between
effective spin-$1/2$ Kondo correlation and superconductivity as in the strong
coupling limit. Since the superconducting gap $(\Delta/T_K^0 \gg 1)$ is larger
than in the strong coupling limit $(\Delta/T_K^0 \ll 1)$, however, the
transition can take place at smaller values of $t/\Gamma$ for which the
effective Kondo temperature $T_K$ is higher.

Moreover, the critical current is relatively large even though the
system is in the weak coupling limit, while in the single quantum dot
the critical current in this limit is very small ($I/I_c^s < 0.1$)
\cite{ChoiMS04a}. A very likely explanation is that the one-electron
spin-$1/2$ Kondo state is formed at smaller values of $t/\Gamma$ and
that strong superconductivity is responsible for it.
The (one-electron) Kondo assisted tunneling then makes the junction
more transparent and enhances the critical current. Hence, the
physical origin of the peak in the critical current is different in
the weak and strong coupling limits. 

\paragraph{Intermediate coupling ($T_K\sim\Delta$).---}

A highly non-trivial behavior occurs for $T_K\sim\Delta$.  In this
regime, the superconductivity, the superexchange, and the Kondo
physics can all be in close competion.  This subtle interplay keeps the
Josephson critical current finite, somewhere between the weak coupling
and strong coupling limit, Fig.~\ref{superdots::fig:1} (c and d),
except for very large $t/\Gamma$, where single-dot physics again
governs the transport.

More interestingly, the phase diagrams in Fig.~\ref{superdots::fig:2} (c,d)
reveal the reentrance behavior from the $0$-junction, to $\pi'$-junction, and
back to $0$-junction, and eventually to the $\pi$-junction for larger
superconducting phase difference $\phi$.  In order to understand this
behavior, we closely examine the subgap Andreev bound states, which are
Bogoliubov quasi-particle excitations from the ground state\cite{Tinkham96a}
and whose derivatives with respect to $\phi$ give the Josephson current
\cite{endnote:2}.  In Fig.~\ref{superdots::fig:3} we plot the energy levels
of the Andreev states as a function of $\phi$, 
in the parameter regime corresponding to the island in
Fig.~\ref{superdots::fig:2} (c).  Let us focus on, say, $\phi=0.9\pi$.
For $t/\Gamma<0.1$, the singlet Kondo clouds are formed between the
superconductors and the adjacent dots, thus the ground state is
likewise a spin singlet, while the excitations correspond to doublet
states, see Fig.~\ref{superdots::fig:3} (b). As the Josephson current
is given approximately by the phase-difference-derivative of the
Andreev levels, this corresponds to the $0$-junction behavior;
Fig.~\ref{superdots::fig:3} (a).
For $0.4<t/\Gamma<5$, the local inter-dot singlet state is induced on
the DQD due to the antiferomagnetic superexchange interaction, thus
the ground state is again a spin singlet; Fig.~\ref{superdots::fig:3}
(h). As before, this results in the $0$-junction behavior;
Fig.~\ref{superdots::fig:3} (g).

In the above two cases, both Kondo effect and superexchange barely win
over the superconductivity, for all values of $\phi$.  However, when
$0.1<t/\Gamma<0.4$, the Kondo effect is suppressed by the large phase
difference.\cite{ChoiMS04a} This is indicated in the fact that the
ground state is now a doublet, while the excited state is a singlet,
as shown in Fig.~\ref{superdots::fig:3} (d,f). Accordingly, the
transport properties are different and, in particular, the
$\pi$-junction behavior is observed; see Fig.~\ref{superdots::fig:3}
(c,e). This regime is denoted as $\pi'$ in the phase diagram in
Fig.~\ref{superdots::fig:2}. While the $\pi$ phase for large $t$
corresponds to the single occupancy of the dots, the $\pi'$ phase
occurs for the double occupancy. For large values of $\Gamma$
($\Gamma=0.02D$), the $\pi'$ island becomes bigger and it can merge
with the $\pi$ regime.

\paragraph{Conclusion.---}

We have studied the Josephson current through a serial double quantum dot
coupled to two superconducting leads.
We have observed a strong renormalizations of $t$, the significant
role of the superexchange coupling $J$, and a rich phase diagram
featuring different $0$-$\pi$ transitions. In particular, when both
the superconductivity and the superexchange were in close competion
with the Kondo physics ($\Delta\sim J\sim T_K$), there appeared an
island of $\pi'$-phase at larger values of the superconducting phase
difference. This finding motivates further studies of this regime, which
may shed new light on the physics of heavy-fermion superconductors.

\begin{acknowledgments}
M.-S.C. was supported by the NRF grant (2009-0080453), the BK21 program, and
the APCTP.  R. A. was supported by MEC-Spain (Grant No. MAT2006-03741, FIS2009-08744), CSIC and CAM (Grant No. CCG08- CSIC/MAT- 3775) M.-S.C. and R.A. acknowledge useful discussions with A. Levy Yeyati.
R. Z. was supported by ARRS grant no. Z1-2058.
\end{acknowledgments}


\begin{thebibliography} {99}
\bibitem{endnote:2} In this qualitative discussion we ignore the small
contribution from the continuum of states above the gap.
\end{thebibliography}
\end{document}